\documentclass[twocolumn]{svjour3}
\smartqed
\usepackage{graphicx}
\journalname{Granular Matter}
\begin{document}

\title{Stochastic Generation of Particle Structures with Controlled Degree of Heterogeneity}

\author{Iwan Schenker \and Frank T. Filser \and Ludwig J. Gauckler}

\institute{I. Schenker (corresponding author) \and F. T. Filser
\and L. J. Gauckler \at
              Nonmetallic Materials, Department of Materials, ETH Zurich, Zurich CH-8093, Switzerland\\
              \email{iwan.schenker@alumni.ethz.ch}           
           }

\date{Received: 16. 2. 2010 / Published online: 17. 4. 2010}

\sloppy

\maketitle

\begin{abstract}
The recently developed void expansion method (VEM) allows for an
efficient generation of porous packings of spherical particles
over a wide range of volume fractions. The method is based on a
random placement of the structural particles under addition of
much smaller ``void-particles'' whose radii are repeatedly
increased during the void expansion. Thereby, they rearrange the
structural particles until formation of a dense particle packing
and introduce local heterogeneities in the structure. In this
paper, microstructures with volume fractions between 0.4 and 0.6
produced by VEM are analyzed with respect to their degree of
heterogeneity (DOH). In particular, the influence of the void- to
structural particle number ratio, which constitutes a principal
VEM-parameter, on the DOH is studied. The DOH is quantified using
the pore size distribution, the Voronoi volume distribution and
the density-fluctuation method in conjunction with fit functions
or integral measures. This analysis has revealed that for volume
fractions between 0.4 and 0.55 the void-particle number allows for
a quasi-continuous adjustment of the DOH. Additionally, the
DOH-range of VEM-generated microstructures with a volume fraction
of 0.4 is compared to the range covered by microstructures
generated using previous Brownian dynamics simulations, which
represent the structure of coagulated colloidal suspensions. Both
sets of microstructures cover similarly broad and overlapping
DOH-ranges, which allows concluding that VEM is an efficient
method to stochastically reproduce colloidal microstructures with
varying DOH.

 \keywords{degree of heterogeneity \and discrete element method \and microstructure generation \and porosity \and void expansion method}
\end{abstract}

\section{Introduction}\label{intro}

The mechanical properties of colloidal particle gels are of great
importance in nature and in many of today's technologies.
Sediments~\cite{Yun_2007}, clay~\cite{Touiti_2009}, some
food~\cite{Mezzenga_2005} or paints and
coatings~\cite{Barbesta_2001} are but a few examples. Typically,
colloidal gels exhibit very complex mechanical behaviors such as
shear thickening~\cite{Lee_2003}, thixotropy~\cite{Barnes_1997} or
aging~\cite{Abou_2001}. From a scientific perspective, these
mechanical properties are often studied using random sphere
packings. They provide ubiquitous model systems for colloids -- or
granular materials in general -- and allow for a systematic study
of the mechanical properties as a function of the various
parameters as, for example, the volume
fraction~\cite{Zaccone_2007}, the particle size
distribution~\cite{Gardiner_2006}, material properties such as the
particles' friction coefficients~\cite{Silbert_2002} or adhesive
forces~\cite{Martin_2008}.

Our present research focuses on the influence of the
microstructural arrangement of the primary particles on the
structure's mechanical properties. This influence is often
observed implicitly in experimental and simulated mechanical tests
on gels or particle packings differing in preparation history.
Macroscopic stress profiles, for example, were found to depend
strongly on the sample preparation procedure and thus on the
microstructure~\cite{Atman_2005}. Franks \it et al. \rm
\cite{Franks_2004} have investigated the mechanical properties of
sedimented aggregates of submicron alumina particles and, in
particular, the influence of the aggregate size (i.e. the
microstructure) on the rheological properties. They found that
sediments formed from larger aggregates exhibit higher shear and
compressive yield strengths. In~\cite{Ref1,Ref2,Ref3}, the
influence of the heterogeneity of a coagulated colloidal
suspension on its mechanical properties was investigated. It was
shown that more heterogeneous microstructures exhibit up to ten
times higher elastic properties and yield strengths than their
more homogeneous counterparts at equal volume fraction. These
experiments have systematically shown that the local arrangement
of the powder particles has a strong influence on the mechanical
properties. The relation between the microstructure and the
mechanical properties, however, is still an open question.

A precondition for any methodical study of the relation between
the microstructure of any granular material and its mechanical
properties is the possibility to generate particle packings with a
controlled degree of heterogeneity (DOH). For colloidal particle
structures, this reproducible control of the heterogeneity is
experimentally achieved using an enzyme catalyzed gelation method
(DCC = direct coagulation casting~\cite{Ref4,Ref5}). DCC allows
for an undisturbed coagulation of electrostatically stabilized
colloidal suspensions to stiff microstructures by an \it in situ
\rm transition of the inter-particle potential from repulsive to
attractive. This can be done along two principal pathways:
shifting the pH of the suspension to the particles' isoelectric
point ($\mathrm{\Delta}$pH-method) or increasing the ionic
strength in the suspension at constant pH
($\mathrm{\Delta}$I-method), which compresses the Debye length of
the repulsive inter-particle potential. The first method leads to
more homogeneous microstructures through diffusion-limited
aggregation. The second method produces more heterogeneous
microstructures via reaction-rate-limited aggregation~\cite{Ref2}.

A variation of the $\mathrm{\Delta}$pH-method producing
microstructures with higher degree of heterogeneity consists in
admixing alkali-swellable polymer (ASP) particles to the powder
particles under acidic conditions. The initial diameter of the ASP
particles is small (approximately 80~nm) in comparison to the
diameter of the structural particles (400~nm). The ASP particles
swell upon increasing pH during the internal gelling reaction and
unfold to roughly 700~nm in diameter, thereby pushing the
structural particles in their vicinity, creating larger pores and
thus producing more heterogeneous microstructures.

One way to investigate the microstructure-dependent mechanical
properties of coagulated colloidal structures is by computational
means, such as the discrete element method or molecular dynamics.
These methods take intrinsically account of the colloid's
particulate nature but require to be provided with the initial
particle configurations. In the case of gravitationally stable
random particle packings with volume fractions between random
loose ($\Phi_{RLP} \approx 0.55$~\cite{Ref6b}) and random close
packing ($\Phi_{RCP} \approx 0.64$~\cite{Ref6c}) a large variety
of algorithms exists. In~\cite{Ref6d}, for example, an overview of
available approaches is given. Basically, they can be categorized
into two main branches: dynamic techniques and constructive
algorithms. Dynamic techniques, to which the method used in this
study belongs, rely on a gradual swelling of the particles or
shrinking of the simulation box. Constructive methods, such as the
method introduced in~\cite{Ref6d}, construct densely packed
particle assemblies using geometrical calculations. Volume
fractions below $\Phi_{RLP}$ were achieved by molecular dynamics
simulations using long-range attractive interactions between the
particles and high friction coefficients~\cite{Ref6e}.

Despite the large amount of structure generation methods, to our
knowledge, none of them has been shown to provide the possibility
to control the microstructural arrangement of the particles
independently of other parameters such as the volume fraction or
the particle size distribution. This control, however, is crucial
to our study of the microstructure-dependent mechanical properties
and was a motivation to develop the void expansion method (VEM)
presented in~\cite{Ref7}. The method was originally inspired by
the generation of heterogeneous microstructures using ASP
particles and was shown to allow for a fast and efficient
computational generation of porous particle structures over a
broad range of volume fractions.

In this paper, particle packings with volume fractions between 0.4
and 0.6 generated using the void expansion method are further
analyzed in terms of their degree of heterogeneity (DOH). The DOH,
introduced in~\cite{Ref8}, represents an abstract concept
quantitatively describing the heterogeneity of a particle
arrangement by scalar measures. In order to quantify the DOH,
three structure characterization methods in combination with
parameters in fit functions or integral measures are used: the
pore size distribution, the Voronoi volume distribution and the
density-fluctuation method. For volume fractions between 0.4 and
0.55, we show that the number of void-particles strongly
influences the DOH of the final microstructure. Additionally, for
a volume fraction of 0.4, the degrees of heterogeneity of
structures generated using VEM and previous Brownian dynamics (BD)
simulations~\cite{Ref9,Ref10} are compared. The BD microstructures
represent coagulated colloidal microstructures, for which the DOH
was shown to depend on the presence and depth of a secondary
minimum in the inter-particle potential described by the
Derjaguin-Landau-Verweg-Overbeek (DLVO) theory~\cite{Ref11}. In
contrast to these BD-simulations that simulate the coagulation
using widely accepted physical laws and theories, VEM is a
stochastic method that has the advantage of its computational
efficiency. We found that the DOH-ranges of the VEM- and the
BD-microstructures overlap to a large extent with the range of the
VEM-microstructures shifted to slightly higher values. This
indicates that VEM is well-suited to reproduce coagulated
colloidal microstructures with varying degree of heterogeneity.

\section{Materials and Methods}\label{sec:vem2_2}

\subsection{Void Expansion Method}\label{sec:vem2_21}

The void expansion method is implemented using the particle flow
code in three dimensions (PFC$^{\mathrm{3D}}$) from Itasca
Consulting Group, Inc., Minneapolis, Minnesota, USA~\cite{Ref12}.
PFC$^{\mathrm{3D}}$ is based on the discrete element
method~\cite{Ref13}, which allows a modelling of the movement of
assemblies of rigid spherical particles. In PFC$^{\mathrm{3D}}$, a
central-difference scheme is used to numerically integrate the
accelerations and velocities of the particles and thus to
determine their dynamic behavior. The forces on the particles
included in our model arise from a linear elastic contact law
between the particles and damping. The contact law is
characterized by the particles' normal and shear stiffness $k_n$
and $k_s$, respectively and the damping force is adjusted via the
damping coefficient $d$~\cite{Ref14}.

The void expansion method relies on two kinds of particles:
``structural particles'' and ``void-particles''. The structural
particles, with normal stiffness $k_{n,S}$ and shear stiffness
$k_{s,S}$, constitute the final microstructures, whereas the
void-particles (normal and shear stiffness $k_{n,V}$ and
$k_{s,V}$, respectively) are only used during structure
generation. The edge length $L_{box}$ of the cubic simulation box
with periodic boundary conditions is calculated using the number
of structural particles $N_S$, their radius $r_S$ and their volume
fraction $\Phi_S$ via

\begin{equation}
 L_{box}=r_s\left(\frac{4N_S\pi}{3\Phi_S}\right)^{1/3}.
 \label{eq:vem2_edgelength}
\end{equation}

The $N_S$ structural particles are randomly placed in the
simulation box. In PFC$^{\mathrm{3D}}$, particles are not allowed
to overlap during their generation. Thus, in order to achieve
volume fractions higher than approximately 0.35, the structural
particles are generated with a reduced initial radius of
$r_S/(m+1)$, $m = 10$ in our simulations. The $N_V$ void-particles
having a radius $r_V \ll r_S$ are randomly added to the structural
particles. After generation of all the particles, the radius of
the structural particles is increased by means of $m$ repeated
radius blow-up steps. At each step, the initial particle radius is
added to the current radius, followed by an equilibration of the
structure until, after $m$ steps, the final particle radius $r_S$
is reached.

After the structural particles have reached their final size, the
radius of the void-particles is cyclically increased by adding
their initial radius $r_V$ to their current radius in alternation
with the performance of 20000 calculation steps in order to allow
for a relaxation of the structure. This cyclic increase of the
void-particle radius simulates the swelling of the ASP particles
in the experiment and causes the structural particles to rearrange
and to get in contact with each other.

The iteration is generally done until the structural and
void-particles are densely packed and any further increase in the
void-particle radius leads to a compaction of the structural
particles. This is reflected by an increasing internal strain
energy, which essentially depends on the stiffness of the void-
and the structural particles. In this paper, VEM-microstructures
having a specific average coordination number are chosen for
further analysis. For a volume fraction of 0.4, an average
coordination number of $CN_{0.4} = 4.7$ is targeted. This value
corresponds to the average coordination number of the
microstructures resulting from the aforementioned BD-simulations,
thus allowing for a direct comparison of their degrees of
heterogeneity. For volume fractions above 0.4, the according
coordination numbers are determined using a linear interpolation
between $CN_{0.4}$ and $CN_{RCP}=6$~\cite{Ref14b}. All analyses
are performed on the structural particles alone, after deletion of
the void-particles. As shown in~\cite{Ref7}, the evolution of $CN$
as a function of the void-particle radius depends on the
void-particle number $N_V$. The range of the $N_V$-values that
allow reaching the targeted $CN$ is discussed in
Sec.~\ref{sec:vem2_34} in terms of particle overlaps.

The simulation parameters are summarized in
Table~\ref{tab:vem2_1}. In order to reduce the inertia of the
void-particles their density $\rho_V$ was set ten times smaller
than the density of the structural particles $\rho_S$, for which
the density of bulk alumina was chosen. The inter-particle
friction coefficient $\mu$ was set to zero in order not to impede
any particle rearrangements.

\begin{table}
\caption{Simulation parameters} \label{tab:vem2_1}
\begin{tabular}{lll}
\hline\noalign{\smallskip}
Parameter & Symbol & Value  \\
\noalign{\smallskip}\hline\noalign{\smallskip}
Number of particles & $N_S$ & 8000 \\
Particle radius & $r_S$ & 2.5 $\times$ 10$^{-7}$ m \\
Normal structural particle stiffness & $k_{n,S}$ & 10$^{3}$ N/m \\
Shear structural particle stiffness & $k_{s,S}$ & 10$^{-2}$ N/m \\
Number of void-particles & $N_V$ & 1000 -- 16000 \\
Normal void-particle stiffness & $k_{n,V}$ & 10$^{2}$ N/m \\
Shear void-particle stiffness & $k_{s,V}$ & 10$^{-2}$ N/m \\
Damping coefficient & $d$ & 0.7 \\
Friction coefficient & $\mu$ & 0.0 \\
Volume fraction & $\Phi_S$ & 0.4 -- 0.6 \\
Structural particle density & $\rho_S$ & 3690 kg/m$^{3}$ \\
Void-particle density & $\rho_V$ & 369 kg/m$^{3}$ \\
\noalign{\smallskip}\hline
\end{tabular}
\end{table}

\subsection{Quantification of the Degree of Heterogeneity}\label{sec:vem2_22}

Three distinct structural characterization methods are used to
analyze and assess the heterogeneity of the
VEM-microstructures~\cite{Ref8}: the pore size distribution, the
density-fluctuation method and the distribution of Voronoi
volumes. Each method, in conjunction with fit functions or
integral measures, provides a scalar measure that captures and
quantifies the DOH of a microstructure.

\subsubsection{Pore Size Distribution}\label{sec:vem2_221}

The pore size distribution is calculated using the exclusion
probability $E_V(r)$~\cite{Ref15}. It is defined as the
probability of inserting a ``test'' particle of radius $r$ at some
arbitrary position in the pore space of a microstructure and is
estimated using a Monte Carlo approach. The probability $P(r_P>r)$
of finding a pore with radius $r_P$ larger than $r$ is obtained
using $P(r_P > r) = \sum_{r' > r_P}E_V(r')$. For coagulated
colloidal suspensions resulting from BD-simulations it was shown
that $P(r_P > r)$ follows a complementary error function given by

\begin{equation}
P(r_P > r) = 1 - \mathrm{erf} \left( \frac{r/r_0 -b}{a \sqrt{2}}
\right),
 \label{eq:vem2_compl_error_fxn}
\end{equation}

\noindent with $a$ and $b$ the standard deviation and the mean
value, respectively. Parameter $a$ was shown to nicely reflect the
structure's degree of heterogeneity, where increasing values of
$a$ indicate an increasing DOH.

\subsubsection{Density Fluctuation Method}\label{sec:vem2_222}

The density-fluctuation method statistically analyzes the spatial
distribution of the particle centers as a function of grid
spacing. Therefor, the cubic simulation box is subdivided into
$n_c^3$ cells where $n_c=2,\ldots,n_c^{max}$ ($n_c^{max}=33$,
cf.~\cite{Ref8}). The standard deviation $\sigma_{ppc}$ normalized
by the average particle number $E_{ppc}$ as a function of grid
spacing was shown to depend on the DOH of the microstructure and
the integral $I_{df}$ over these curves (Eq.~(\ref{eq:vem2_Idf}))
provides a measure of the DOH.

\begin{equation}
I_{df} = \sum_{n_c \leq n_c^{max}}
\frac{\sigma_{ppc}(n_c)}{E_{ppc}}
 \label{eq:vem2_Idf}
\end{equation}

\subsubsection{Voronoi Volume Distribution}\label{sec:vem2_223}

The Voronoi volume of a particle is the volume given by the union
of all points in pore space closer to the surface of this particle
than to any other particle~\cite{Ref16}. The Voronoi tessellation
thus divides a set of particles into a set of space-filling,
non-overlapping and convex polyhedrons. In this study, the
polyhedron volumes are determined using the \it Qhull \normalfont
package~\cite{Ref17}. Following a statistical mechanics approach,
Aste and Di Matteo have shown in~\cite{Ref18} that the Voronoi
volume distribution of a particle set follows a so-called
$k$-gamma distribution given by

\begin{equation}
f(V^f, k) = \frac{k^k}{\Gamma(k)}
\frac{(V^f)^{k-1}}{(\bar{V}^f)^k}\exp(-k \frac{V^f}{\bar{V}^f}).
 \label{eq:vem2_kgamma}
\end{equation}

The Voronoi free volume $V^f = V - V_{min}$ is the difference
between a particle's Voronoi volume $V$ and the minimum volume of
a Voronoi cell $V_{min}$, which is achieved for a regular close
packing and is given by $V_{min} = 1.325\,V_{sphere}$, with
$V_{sphere}$ the volume of a particle. The mean Voronoi free
volume $\bar{V}^f$ is a scaling parameter and the free parameter
$k$ characterizes the shape of the curve that very sensitively
depends on the structural organization of the particles.
In~\cite{Ref8}, the $k$-gamma distribution was used to fit the
Voronoi volume distribution of BD-microstructures. A very good
agreement between data and fit was obtained and it was shown that
parameter $k$ reflects very sensitively the DOH of the various
microstructures.

\section{Results and Discussion}\label{sec:vem2_3}

\begin{figure}
  \includegraphics[width=\linewidth]{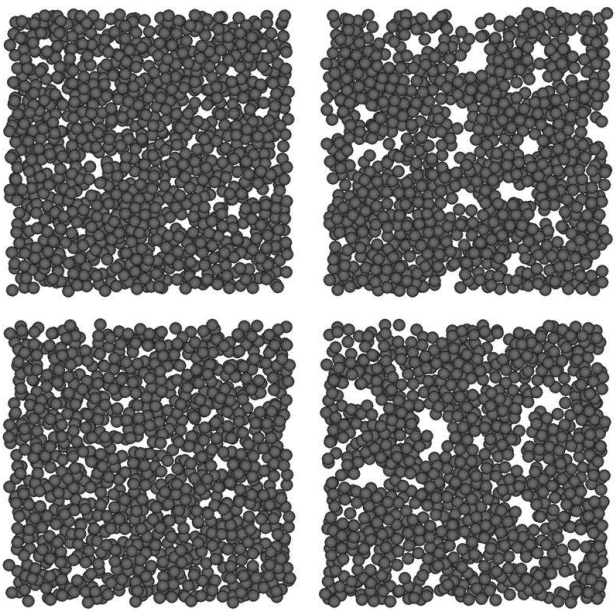}
\caption{Slices through VEM-microstructures (upper row) with a
more homogeneous microstructure ($N_V=16000$, left) and a more
heterogeneous microstructure ($N_V=1000$, right). The lower row
shows the most homogeneous ($\Psi_0=0$~mV, left) and most
heterogeneous microstructure ($\Psi_0=15$~mV, right) resulting
from earlier BD-simulations~\cite{Ref9}. Slice thickness: three
particle diameters; particle diameter~=~0.5~$\mu$m, volume
fraction~=~0.4 (for all structures).} \label{fig:vem2_fig_1}
\end{figure}

A qualitative impression of the influence of the void-particle
number $N_V$ on the resulting microstructures is given in
Fig.~\ref{fig:vem2_fig_1}, presenting slices with a thickness of
three particle layers and equal volume fraction of 0.4. In the
upper row, slices through VEM-microstructures generated using
$N_V=16000$ (left) and $N_V=1000$ (right), respectively, are
shown. In the left slice, the particles are rather uniformly
distributed, whereas the particles in the slice on the right are
locally more densely packed and thus larger voids can be observed.
Very similar properties are observed for the BD-microstructures
inserted in the lower row. On the left, the most homogeneous
(surface potential $\Psi_0=0$~mV) and, on the right, the most
heterogeneous BD-microstructure ($\Psi_0=15$~mV) are shown.
Figure~\ref{fig:vem2_fig_1} suggests that the use of larger
void-particle numbers leads to more homogeneous microstructures
and that VEM allows obtaining very similar microstructures as BD
in terms of heterogeneity, as supported by the quantitative
analyses in Sec.~\ref{sec:vem2_31}. In Sec.~\ref{sec:vem2_34}, the
DOH of the microstructures generated using VEM and BD are compared
and, in Sec.~\ref{sec:vem2_35}, the possibility to generate
microstructures with a varying DOH for volume fraction above 0.4
is investigated.

\subsection{Influence of the Void-Particle Number}\label{sec:vem2_31}

In this section, the results obtained for the three distinct
methods used to quantify the DOH are presented in detail for the
VEM-microstructures with a volume fraction of 0.4 and varying
void-particle number $N_V$.

\subsubsection{Pore Size Distribution}\label{sec:vem2_311}

\begin{figure}
  \includegraphics[width=\linewidth]{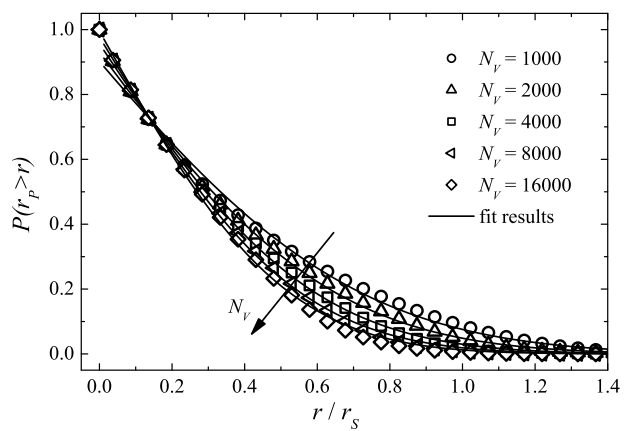}
\caption{Probability $P(r_P > r)$ of finding pores with a radius
$r_P$ larger than $r$ vs.~$r/r_S$ obtained using the pore size
distribution for the various VEM-microstructures with $\Phi_S=0.4$
(symbols). Solid lines denote the corresponding fits using a
complementary error function (Eq.~\ref{eq:vem2_compl_error_fxn}).}
\label{fig:vem2_fig_2}
\end{figure}

The probability $P(r_P > r)$ of finding pores with a radius $r_P$
larger than $r$ is shown in Fig.~\ref{fig:vem2_fig_2} for various
VEM-microstructures with void-particle numbers ranging from 1000
to 16000 (symbols). The data is shown as a function of $r$
normalized by the particle radius $r_S$. For a given pore radius
$r_P > 0$, $P(r_P > r)$ decreases for increasing void-particle
numbers $N_V$. The probability of finding pores with a radius
larger than $0.5\,r_S$ is 1.6 times higher in the $N_V=1000$ than
in the $N_V=16000$ microstructure. Finding pores larger than
$0.75\,r_S$ and $1.0\,r_S$ is 4.3 and 18.7 times more probable in
the microstructure with $N_V=1000$ than in the one with
$N_V=16000$, respectively. The solid lines in
Fig.~\ref{fig:vem2_fig_2} represent the complementary error
function fit curves obtained using
Eq.~(\ref{eq:vem2_compl_error_fxn}). Fit parameters $a$ and $b$
and  the  corresponding  $R^2_{a,b}$-values are summarized in
Table~\ref{tab:vem2_2}. The $R^2_{a,b}$-values close to one
indicate excellent fits. The largest value $a$ is found for the
microstructure generated using 1000 void-particles. For increasing
values of $N_V$, parameter $a$ decreases, which reflects a
decreasing probability of finding larger pores and thus a
decreasing DOH.

\begin{table}
\caption{Measures of the degree of heterogeneity as a function of
the void-particle number $N_V$: fit parameters $a$ and $k$,
corresponding to the widths of the pore size and Voronoi volume
distribution, respectively, and the integral over the density
fluctuation curves $I_{df}$. Additionally, the second fit
parameter of the cumulative pore size distribution ($b$) and the
corresponding $R^2$-values are shown.} \label{tab:vem2_2}
\begin{tabular}{lllllll}
\hline\noalign{\smallskip}
$N_V$ & $a$ & $b$ ($10^{-2}$) & $R^2_{a,b}$ & $I_{df}$ & $k$ & $R^2_k$ \\
\noalign{\smallskip}\hline\noalign{\smallskip}
  1000 & 0.605 & -7.44  & 0.997 & 24.2 & 2.46 & 0.977\\
  2000 & 0.530 & -4.63  & 0.999 & 23.6 & 2.99 & 0.995\\
  4000 & 0.472 & -2.56  & 0.999 & 23.1 & 3.61 & 0.998\\
  8000 & 0.431 & -1.21  & 0.999 & 22.7 & 4.67 & 0.999\\
 11000 & 0.416 & -0.652 & 0.999 & 22.6 & 5.02 & 0.999\\
 13000 & 0.411 & -0.458 & 0.998 & 22.6 & 5.50 & 0.999\\
 16000 & 0.401 & -0.0227 & 0.999 & 22.5 & 5.74 & 0.998\\
\noalign{\smallskip}\hline
\end{tabular}
\end{table}

\subsubsection{Density Fluctuation Method}\label{sec:vem2_312}

\begin{figure}
  \includegraphics[width=\linewidth]{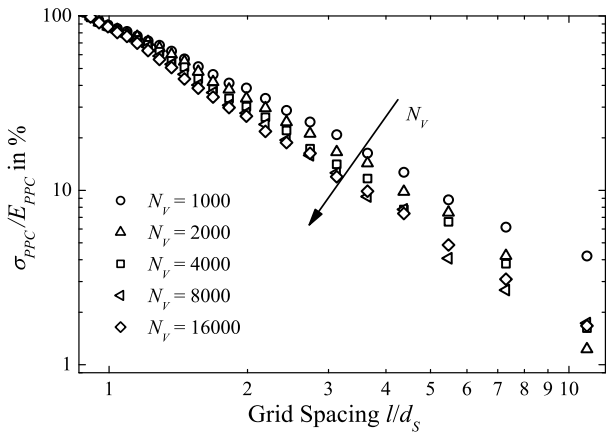}
\caption{Density fluctuations as a function of grid spacing for
the various VEM-microstructures with $\Phi_S=0.4$.}
\label{fig:vem2_fig_3}
\end{figure}

The density fluctuations of the various VEM-microstructures are
presented in Fig.~\ref{fig:vem2_fig_3} as a function of the grid
spacing normalized by the particle diameter $d_S = 2\,r_S$. The
density fluctuations decrease with increasing void-particle
number. This behavior is quantified using $I_{df}$ as given in
Eq.~(\ref{eq:vem2_Idf}). The values are summarized in
Table~\ref{tab:vem2_2}. $I_{df}$ increases with decreasing $N_V$
and thereby reflects an increasingly heterogeneous repartition of
the particles.

\subsubsection{Voronoi Volume Distribution}\label{sec:vem2_313}

\begin{figure}
  \includegraphics[width=\linewidth]{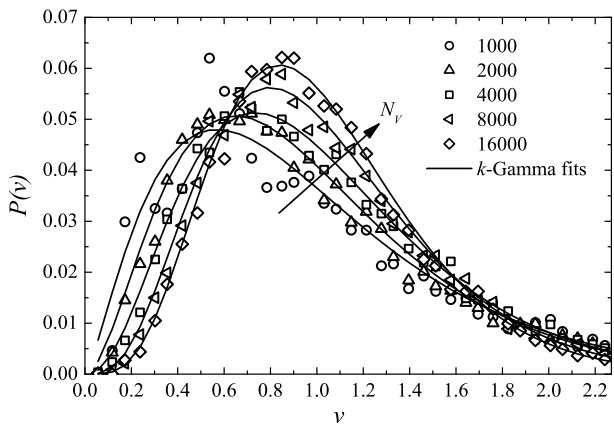}
\caption{Voronoi volume distribution $P(\nu)$ (symbols) as a
function of $\nu$ and corresponding $k$-gamma fits (lines) for
various VEM-microstructures ($\Phi_S=0.4$).}
\label{fig:vem2_fig_4}
\end{figure}

The Voronoi volume distribution $P(\nu)$ of the various
VEM-microstructures is shown in Fig.~\ref{fig:vem2_fig_4} for
void-particle numbers ranging between 1000 and 16000 (symbols) as
a function of $\nu = \frac{V^f}{\bar{V}^f}$, the free Voronoi
volume normalized by the mean free volume. For increasing $N_V$,
the peak height increases and, consequently, the width of the
curve decreases. This behavior indicates that more homogeneous
microstructures are found toward increasing $N_V$, which is
confirmed by a fit of the curves using the $k$-gamma distribution
given in Eq.~(\ref{eq:vem2_kgamma}). The fits are shown as lines
in Fig.~\ref{fig:vem2_fig_4}. The corresponding values for the
parameter $k$, summarized in Table~\ref{tab:vem2_2}, increase for
increasing $N_V$ reflecting the shift toward more homogeneous
microstructures. Very good fit results are achieved as indicated
by the $R^2_k$-values.

\subsection{Degree of Heterogeneity for VEM- and BD-Structures at volume fraction~0.4}\label{sec:vem2_34}

\begin{figure}
  \includegraphics[width=\linewidth]{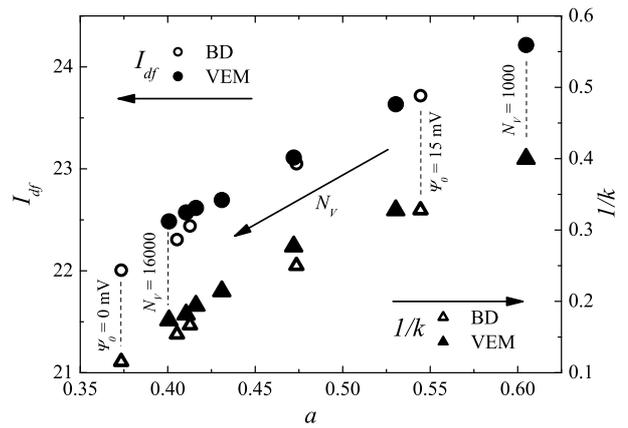}
\caption{Interdependence of the DOH-measures ($\Phi_S=0.4$):
Integral over the density fluctuation curves $I_{df}$ (circles,
left scale) and width of the Voronoi volume distribution $1/k$
(triangles, right scale) as a function of the width of the pore
size distribution $a$ for the various VEM-microstructures (full
symbols) and for the BD-microstructures (open symbols~\cite{Ref8})
at $\Phi_S=0.4$.} \label{fig:vem2_fig_5}
\end{figure}

The interdependence between the three measures of the degree of
heterogeneity is shown in Fig.~\ref{fig:vem2_fig_5}. On the left
scale, the integral over the density fluctuation curves $I_{df}$
is shown as a function of the width of the pore size distribution,
given by parameter $a$. The corresponding values are shown as
circles. The right scale presents $1/k$ reflecting the width of
the Voronoi volume distribution in dependence of $a$, shown as
triangles. Full symbols denote the values obtained for the
VEM-microstructures.

The linear arrangement of the two curves suggests a pairwise
affine relation between the various DOH. Furthermore, the values
found for the VEM-microstructures are very close to those obtained
for the BD-microstructures, which are inserted as open symbols.
These two points substantiate the equivalence of the three methods
as already observed in~\cite{Ref8}.

The DOH-range covered by the VEM-microstructures, in terms of
parameter $a$, extends from 0.4 to 0.6. This overlaps to a large
extent with the range of the BD-microstructures from 0.37 to 0.54.
Thus, the range covered by VEM is roughly 20\%~broader starting at
a slightly higher DOH than the range of the BD-microstructures
considered in this work.

\begin{figure}
  \includegraphics[width=\linewidth]{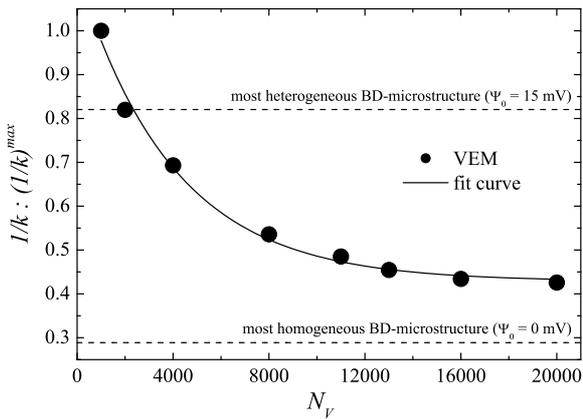}
\caption{DOH-measure $1/k$ relative to $(1/k)^{max}$ as a function
of the void-particle number $N_V$ for $\Phi_S=0.4$ (symbols). The
dashed lines denote the corresponding values of the most and least
heterogeneous BD-microstructure~\cite{Ref8}.}
\label{fig:vem2_fig_6}
\end{figure}

The sensitivity of the DOH on $N_V$ is demonstrated in
Fig.~\ref{fig:vem2_fig_6} presenting $1/k$ relative to its maximum
value $(1/k)^{max}$, achieved for $N_V = 1000$, as a function of
$N_V$. After a strong initial decrease for small values of $N_V$,
$1/k$ levels off toward higher values of $N_V$. A further increase
of $N_V$ has only a negligible influence on the DOH as shown by
the insertion of the additional $1/k$-value for $N_V = 20000$.
Indeed, an increase in $N_V$ from 16000 to 20000 decreases the DOH
by only 2\%, which suggests that the DOH of the most homogeneous
BD-microstructure, represented by the lower dashed line in
Fig.~\ref{fig:vem2_fig_6}, cannot be reached using VEM. This is
supported by the fit of the data using

\begin{equation}
\left(1/k - 1/k_0\right) \propto \exp\left(\beta N_V\right),
 \label{eq:k_N_V}
\end{equation}

\noindent yielding a minimum DOH of $1/k_0 = 0.43$, which is
larger than the $1/k$-value of the homogeneous BD-microstructure.
The second fit parameter and the corresponding correlation
coefficient are $\beta = -0.25 \times 10^{-3}$ and $R^2$ = 0.99,
respectively. The reason for this asymptotic behavior is most
probably related to the VEM-algorithm itself in combination with
the chosen targeted mean coordination number $CN$ = 4.7. As shown
in~\cite{Ref7}, $CN$ as a function of the void-particle radius
follows a characteristic step-like shape with $CN_i$ the
coordination number at the curve's inflection point. $CN_i$ was
found to decrease for increasing void-particle numbers $N_V$.
Above the inflection point, $CN$ scales as a power law with
exponent 0.37, independently of $N_V$. Therefore it becomes
increasingly difficult to reach a targeted $CN > CN_i$ toward
larger values of $N_V$, at least without accepting considerable
particle overlaps and consequently high internal stresses. The
mean particle overlap, as determined by the maximum in the
pair-correlation function was found to be 1.4\% of the particle
diameter $d_S$ in the case of $N_V=13000$ and $CN=4.7$. For
$N_V=16000$ the mean overlap is 1.5\%~$d_S$, however with a
slightly lower $CN$ of 4.63. The maximum particle overlap is below
2\%~$d_S$ for all microstructure with $N_V \leq 8000$ and
increases to roughly 4\%~$d_S$ for $N_V=16000$. For $N_V = 20000$
a mean and maximum overlap of 2\%~$d_S$ and 5\%~$d_S$,
respectively was already found for $CN = 4.57$, which led to the
conclusion that $N_V = 16000$ and thus a void- to structural
particle number ration $n^{max} = 2.0$ constitutes the upper limit
of VEM, for a volume fraction of $\Phi_S = 0.4$.

The lower limit of $N_V$, at around 1000, is mainly imposed by the
decreasing fit quality obtained in the case of the pore size and
Voronoi volume distribution, as expressed in terms of the
$R^2$-values (Table~\ref{tab:vem2_2}). The minimum void- to
structural particle number ratio in the case of $\Phi_S = 0.4$ is
thus $n^{min} = 0.125$.

\subsection{Influence of the Volume Fraction on the DOH}\label{sec:vem2_35}

In the following, the DOH-range is investigated for volume
fractions above 0.4. In particular, VEM-microstructures with
volume fractions $\Phi_S$~=~0.45, 0.50, 0.55 and 0.60 have been
generated and analyzed analogously to the structures with
$\Phi_S=0.4$. This range of $\Phi_S$ was chosen since it is of
particular interest for further simulations of the
microstructure-dependent properties of coagulated colloidal
suspensions~\cite{Ref3}. As mentioned in Sec.~\ref{sec:vem2_21},
the targeted average coordination numbers for the structures with
$\Phi_S > 0.4$ are obtained using a linear interpolation between
$CN_{0.4}=4.7$ and $CN_{RCP}=6.0$ yielding approximately 5.0,
5.25, 5.5 and 5.75, respectively.

\begin{figure}
  \includegraphics[width=\linewidth]{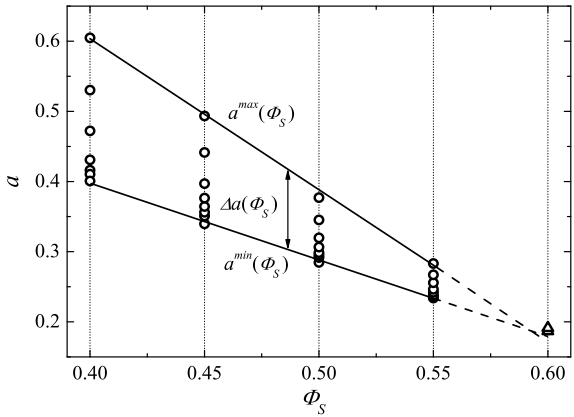}
\caption{DOH-range for volume fractions $\Phi_S$ between 0.4 and
0.6 as expressed by measure $a$ obtained from the pore size
distribution.} \label{fig:vem2_fig_7}
\end{figure}

In this section, the DOH is measured by means of parameter $a$
obtained by fitting the cumulative pore size distribution using
the complementary error function given in
Eq.~\ref{eq:vem2_compl_error_fxn}. This method was chosen because
it yielded the best fit results over the whole range of volume
fractions above 0.4 ($R^2>0.999$ for all fits). The results are
compiled in Table~\ref{tab:vem2_3}. For each volume fraction, an
increasing void-particle number entails a decreasing parameter $a$
and therefore a decreasing DOH. This is summarized in
Fig.~\ref{fig:vem2_fig_7} presenting the DOH-values of these
structures, as measured by parameter $a$ (circles). Particular
interest is given to the maximum and minimum values of $a$ as a
function of $\Phi_S$: $a^{max}(\Phi_S)$ and $a^{min}(\Phi_S)$,
respectively. Our results suggest linear dependencies as
illustrated by the solid lines in Fig.~\ref{fig:vem2_fig_7}. The
respective fits yield:

\begin{equation}
\begin{array}{l}
 a^{max}(\Phi_S) = -2.16\,\Phi_S + 1.47\\
 a^{min}(\Phi_S) = -1.11\,\Phi_S + 0.842.
\end{array}
 \label{eq:a_max_min}
\end{equation}

\noindent which, for the DOH-range measured by parameter $a$,
results in

\begin{equation}
 \Delta a(\Phi_S) = -1.05\,\Phi_S + 0.624.
 \label{eq:delta_a}
\end{equation}

\noindent Equation~\ref{eq:delta_a} describes the narrowing of the
DOH-range toward increasing volume fraction. With respect to the
DOH-range at $\Phi_S=0.4$, the ranges for $\Phi_S=0.45$, 0.5 and
0.55 are reduced by 30, 55 and 76\%, respectively. The linear
extrapolations of $a^{max}(\Phi_S)$ and $a^{min}(\Phi_S)$ toward
higher volume fractions are shown as dashed lines in
Fig.~\ref{fig:vem2_fig_7}. The lines intersect at a volume
fraction of 0.594 with corresponding DOH-value $a_0=0.183$.

\begin{table}
\caption{DOH-measure $a$ as a function of the void-particle number
$N_V$ for various volume fractions $\Phi_S>0.4$. $R^2>0.999$ for
all fits.} \label{tab:vem2_3}
\begin{tabular}{lllll}
\hline\noalign{\smallskip}
$N_V$ & $\Phi_S = 0.45$ & $\Phi_S = 0.50$ & $\Phi_S = 0.55$ & $\Phi_S = 0.60$ \\
\noalign{\smallskip}\hline\noalign{\smallskip}
  500   & -         & -         & 0.283     & 0.191\\
  1000  & 0.493     & 0.377     & 0.267     & -\\
  2000  & 0.441     & 0.345     & 0.256     & -\\
  4000  & 0.397     & 0.320     & 0.247     & -\\
  6000  & 0.376     & 0.307     & 0.242     & -\\
  8000  & 0.364     & 0.299     & 0.238     & 0.187\\
 10000  & 0.356     & 0.295     & 0.236     & -\\
 12000  & 0.351     & 0.292     & 0.234     & -\\
 16000  & 0.340     & 0.285     & -         & -\\
\noalign{\smallskip}\hline
\end{tabular}
\end{table}

The DOH-range for a random packing of monosized spheres is thus
reduced to a unique value at this volume fraction. This was
confirmed using two additional VEM-microstructure at $\Phi_S=0.6$
and using 500 and 8000 void particles (triangles in
Fig.~\ref{fig:vem2_fig_7}). The DOH-measure $a$ for these two
structures yields 0.187 and 0.191, respectively. The range is thus
narrowed down to roughly 2\% with respect to the DOH-range at
$\Phi_S=0.4$.

The decreasing DOH-range toward larger volume fractions is
qualitatively explained by the decrease of the pore space volume
required for particle rearrangements. The volume fraction of 0.6,
however, is below the RCP-limit, which was assumed to be the
volume fraction, at which variations in DOH become zero.
In~\cite{Ref19}, for example, the nature of the RCP limit was
studied by means of a statistical mechanics approach and it was
found that the entropy and thus the number of accessible
configurations of a disordered packing of monosized spheres
reaches a minimum at the RCP limit. Additionally, a sharp decrease
in parameter $k$ was found as the volume fraction crosses the RLP
limit in~\cite{Ref20}. Following our interpretation of $k$, this
sharp decrease corresponds to a sharp increase in DOH, which is
not found in our data. More detailed studies are required in order
to investigate these differences.

\section{Summary and Conclusions}\label{sec:vem2_4}

In this paper, the degree of heterogeneity of microstructures
generated using the void expansion method has been analyzed and
quantified using three distinct techniques based on the pore size
distribution, the density-fluctuation method and the Voronoi
volume distribution. In particular, the influence of the
void-particle number and the volume fraction on the DOH was
investigated.

The various pore size distributions were fitted using a
complementary error function achieving very good fit results over
the whole range of volume fractions investigated in this study.
For all volume fractions, parameter $a$, corresponding to the
width of the pore size distribution, increases for decreasing
$N_V$, thereby reflecting the transition toward a higher DOH. For
this range of volume fractions, the void expansion method thus
allows for a generation of particle arrangements, for which the
DOH can be controlled quasi-continuously over a broad range by
means of the void-particle number. In particular, the DOH-range
becomes narrower and shifts to lower values as a function of
increasing volume fraction. Our results indicate that the
DOH-range is reduced to a single value at a volume fraction of
roughly 0.6, which, interestingly, is below the RCP limit. Further
studies are required in order to analyze the behavior of the DOH
and its range for volume fractions toward and beyond the RCP
limit.

For VEM-structures with volume fraction~0.4, the significant
influence of the void-particle number on the DOH as measured by
parameter $a$ is confirmed by the density-fluctuation method and
the Voronoi volume distribution. $I_{df}$, given by the integral
over the various density fluctuation curves, increases for
decreasing $N_V$. Parameter $k$, characterizing the shape of the
Voronoi volume distributions, decreases for decreasing $N_V$.

The degree of heterogeneity of the VEM-microstructures with a
volume fraction of 0.4 has been compared to that of
BD-microstructures~\cite{Ref9,Ref10} analyzed in~\cite{Ref8}. The
structures generated using BD-simulations represent coagulated
colloidal particle structures where the coagulation was simulated
using widely accepted physical laws and theories. The degree of
heterogeneity of the BD-microstructures was shown to be closely
related to the presence and depth of a secondary minimum in the
DLVO potential, which is determined by the particles' surface
potential. In contrast to BD, VEM is a stochastic method that
mimics the swelling of ASP particles in experiment, having the
advantage of its computational effectiveness. The DOH-range
covered by VEM in terms of parameter $a$, characterizing the width
of the pore size distribution, is approximately 20\%~broader and
shifted to slightly higher values than the DOH-range of these
BD-microstructures. The most homogeneous BD-microstructure, for
which a surface potential $\Psi_0=0$~mV was used, could not be
reproduced by VEM in terms of heterogeneity. A further decrease of
the DOH through an increase in $N_V$ is mostly impeded by the
choice of the targeted coordination number (roughly~4.7). For
void-particle numbers exceeding 16000, this value cannot be
attained without considerable particle overlap and therefore
internal stress in the microstructure. On the other hand, VEM
allows generating more heterogeneous microstructures than these
BD-simulations, where the increasing depth of the secondary
minimum toward higher surface potentials inhibits a complete
coagulation.

This study has undoubtedly shown, that the void-particle number
constitutes an important simulation parameter of VEM with regard
to the DOH of the final microstructure. This parameter indeed
allows for a quasi-continuous control of the DOH over a broad
range of volume fractions. In order to compare the structural
arrangement and to assess the influence of the void-particle
number on the final microstructures, the concept of the DOH itself
has proven particularly useful. The comparison of VEM- and
BD-generated structures at a volume fraction of 0.4 allows
concluding that VEM nicely reproduces the microstructures of
coagulated colloids. Thereby, the void expansion method
facilitates the further study of the microstructure-dependent
mechanical properties of coagulate colloidal structures or
granular matter in general.


\begin{thebibliography}{}

\bibitem{Yun_2007}
Yun, T. S., Santamarina, J. C. and Ruppel, C., Mechanical
properties of sand, silt and clay containing tetrahydrofuran
hydrate, J. Geophys. Res. \bf112\rm, B04106 (2007)

\bibitem{Touiti_2009}
Touiti, L., Bouassida, M. and Van~Impe, W., Discussion on Tunis
Soft Soil Sensitivity, Geotech. Geol. Eng. \bf27 \rm [5], 631--643
(2009)

\bibitem{Mezzenga_2005}
Mezzenga, R., Schurtenberger, P., Burbidge, A. and Michel, M.,
Understanding foods as soft materials, Nature Mater. \bf4\rm,
729-740 (2005)

\bibitem{Barbesta_2001}
Barbesta, F., Bousfield, D. W. and Rigdahl, M., Modeling of
rheological properties of coating colors, J. Rheol. \bf45 \rm [1],
139--160 (2001)

\bibitem{Lee_2003}
Lee, Y. S. and Wagner, N. J., Dynamic properties of shear
thickening colloidal suspensions, Rheol. Acta \bf42 \rm [3],
199--208 (2003)

\bibitem{Barnes_1997}
Barnes, H. A., Thixotropy -- a review, J. Non-Newtonian Fluid
Mech. \bf70\rm, 1--33 (1997)

\bibitem{Abou_2001}
Abou, B., Bonn, D. and Meunier, J., Aging dynamics in a colloidal
glass, Phys. Rev. E \bf64 \rm [2], 021510 (2001)

\bibitem{Zaccone_2007}
Zaccone, A., Lattuada, M., Wu, H. and Morbidelli, M., Theoretical
Elastic Moduli for Disordered Packings of Interconnected Spheres,
J. Chem. Phys. \bf127 \rm [17], 174512 (2007)

\bibitem{Gardiner_2006}
Gardiner, B. S. and Tordesillas, A., Effect of Particle Size
Distribution in a Three-Dimensional Micropolar Continuum Model of
Granular Media, Powder Technol. \bf161 \rm [2], 110-121 (2006)

\bibitem{Silbert_2002}
Silbert, L. E., Erta\c{s}, D., Grest, G. S., Halsey, T. C. and
Levine, D., Geometry of Frictionless and Frictional Sphere
Packings, Phys. Rev. E \bf65 \rm [3], 031304 (2002)

\bibitem{Martin_2008}
Martin, C. L. and Bordia, R. K., Influence of adhesion and
friction on the geometry of packings of spherical particles, Phys.
Rev. E \bf77 \rm [3], 031307 (2008)

\bibitem{Atman_2005}
Atman, A. P. F., Brunet, P., Geng, J., Reydellet, G., Combe, G.,
Claudin, P., Behringer, R. P. and Cl{\'{e}}ment, E., Sensitivity
of the Stress Response Function to Packing Preparation, J. Phys.:
Condens. Matter \bf17\rm, S2391--S2403 (2005)

\bibitem{Franks_2004}
Franks, G. V., Zhou, Y., Yan, Y., Jameson, G. and Biggs, S.,
Effect of Aggregate Size on Sediment Bed Rheological Properties,
Phys. Chem. Chem. Phys. \bf6 \rm [18], 4490--4498 (2004)

\bibitem{Ref1}
Wyss, H. M., Tervoort, E. V. and Gauckler, L. J., Mechanics and
Microstructures of Concentrated Particle Gels, J. Am. Ceram. Soc.
\bf88 \rm [9], 2337--2348 (2005)

\bibitem{Ref2}
Wyss, H. M., Tervoort, E., Meier, L. P., M\"uller, M. and
Gauckler, L. J., Relation between microstructure and mechanical
behavior of concentrated silica gels, J. Colloid Interface Sci.
\bf273 \rm [2], 455--462 (2004)

\bibitem{Ref3}
Wyss, H. M., Deliormanli, A. M., Tervoort, E., and Gauckler, L.
J., Influence of Microstructure on the Rheological Behavior of
Dense Particle Gels, AIChE J. \bf51 \rm [1], 134--141 (2005)

\bibitem{Ref4}
Gauckler, L. J., Graule, Th., Baader, F., Ceramic forming using
enzyme catalyzed reactions, Mater. Chem. Phys. \bf61 \rm [1],
78--102 (1999)

\bibitem{Ref5}
Tervoort, E., Tervoort, T. A. and Gauckler, L. J., Chemical
Aspects of Direct Coagulation Casting of Alumina Suspensions, J.
Am. Ceram. Soc. \bf87 \rm [8], 1530--1535 (2004)

\bibitem{Ref6}
Hesselbarth, D., Tervoort, E., Urban, C. and Gauckler, L. J.,
Mechanical Properties of Coagulated Wet Particle Networks with
Alkali-Swellable Thickeners, J. Am. Ceram. Soc. \bf84 \rm [8],
1689--1695 (2001)

\bibitem{Ref6b}
Agnolin, I. and Roux, J.-N., Internal States of Model Isotropic
Granular Packings. I. Assembling Process, Geometry and Contact
Networks, Phys. Rev. E \bf76 \rm [1], 061302 (2007)

\bibitem{Ref6c}
Jerkins, M., Schr{\"{o}}ter, M., Swinney, H. L., Senden, T. J.,
Saadatfar, M. and Aste, T., Onset of Mechanical Stability in
Random Packings of Frictional Spheres, Phys. Rev. Lett. \bf101 \rm
[1], 018301 (2008)

\bibitem{Ref6d}
Bagi, K., An algorithm to generate random dense arrangements for
discrete element simulations of granular assemblies, Gran. Mat.
\bf7 \rm [1], 31--43 (2005)

\bibitem{Ref6e}
Luding, S., Contact models for very loose granular materials, In:
Eberhard P. (ed) Symposium on Multiscale Problems in Multibody
System Contacts, Springer, Heidelberg, ISBN 978-1-4020-5980-3,
135--150 (2007)

\bibitem{Ref7}
Schenker, I., Filser, F. T., Herrmann, H. J. and Gauckler, L. J.,
Generation of Porous Particle Structures using the Void Expansion
Method, Gran. Mat., \bf11 \rm [3], 201--208 (2009)

\bibitem{Ref8}
Schenker, I., Filser, F. T., Aste, T., Herrmann, H. J. and
Gauckler, L. J., Quantification of the Heterogeneity of Particle
Packings, Phys. Rev. E \bf80 \rm [2], 021302 (2009)

\bibitem{Ref9}
H\"utter, M., Local Structure Evolution in Particle Network
Formation Studied by Brownian Dynamics Simulation, J. Colloid
Interface Sci. \bf231 \rm [2], 337--150 (2000)

\bibitem{Ref10}
H\"utter M., Brownian dynamics simulation of stable and of
coagulating colloids in aqueous suspension, Ph.D. thesis no.
13107, ETH Zurich, Switzerland (1999)

\bibitem{Ref11}
Russel, W. B., Saville, D. A. and Schowalter, W. R., Colloidal
Dispersions, Cambridge University Press (March 1989)

\bibitem{Ref12}
PFC$^{\mathrm{3D}}$ User's Manual, Itasca Consulting Group, Inc.,
Minneapolis, Minnesota, USA (1995)

\bibitem{Ref13}
Cundall, P. A. and Strack, O. D. L., A discrete numerical model
for granular assemblies, G\'eotechnique \bf29 \rm [1], 47--65
(1979)

\bibitem{Ref14}
Brown, E. T., Analytical and Computational Methods in Engineering
Rock Mechanics, Ed. London: Allen \& Unwin (1987)

\bibitem{Ref14b}
Song, C., Wang, P. and Makse, H. A., A phase diagram for jammed
matter, Nature, \bf453 \rm [29], 629--632 (2008)

\bibitem{Ref15}
Torquato, S., Lu, B. and Rubinstein, J., Nearest-Neighbor
Distribution Functions in Many-Body Systems, Phys. Rev. A \bf41
\rm [4], 2059--2075 (1990)

\bibitem{Ref16}
Voronoi, G., Recherches sur les parall{\'{e}}lo{\`{e}}dres
primitives, J. Reine Angew. Math. \bf134\rm, 198--287 (1908)

\bibitem{Ref17}
Barber, C. B., Dobkin, D. P. and Huhdanpaa, H., The Quickhull
Algorithm for Convex Hulls, ACM T. Math. Software \bf22 \rm [4],
469--483 (1996)

\bibitem{Ref18}
Aste, T. and Di Matteo, T., Emergence of Gamma Distributions in
Granular Materials and Packing Models, Phys. Rev. E \bf77 \rm [2],
021309 (2008)

\bibitem{Ref19}
Anikeenko, A. V., Medvedev, N. N. and Aste, T., Structural and
entropic insights into the nature of the random-close-packing
limit, Phys. Rev. E \bf77 \rm [3], 031101 (2008)

\bibitem{Ref20}
Aste, T. and Di Matteo, T., Structural transitions in granular
packs: statistical mechanics and statistical geometry
investigations, Eur. Phys. J. B \bf64 \rm, 511–517 (2008)

\end{thebibliography}
\end{document}